# Effects of Hawking radiation and Wigner rotation on fermion entanglement

Doyeol Ahn*

*Institute of Quantum Information Processing and Systems, University of Seoul, Seoul 130-743, Republic of Korea*

**-Abstract:** In this work, we report that the Hawking radiation effect on fermions is fundamentally different from the case of scalar particles. Intrinsic properties of fermions (exclusion principle and spin) affect strongly the interaction of fermions with both Hawking radiation and metric of the spacetime. In particular we have found the following: first, while the fermion vacuum state seen by the Rindler observer is an entangled state in which the right and left Rindler wedge states appear in correlated pairs as in the case of the scalar particles, the entanglement disappears in the excited state due to the exclusion principle; second, the spin of the fermion experiences the Winger rotation under a uniform acceleration; and third, the quantum information of fermions encoded in spin (entangled state is composed of different spin states but with the same mode function) is dissipated not by the Hawking radiation but by the Wigner rotation as the pair approaches the event horizon.

E-mail: dahn@uos.ac.kr, davidahn@hitel.net

Hawking radiation [1-4] (quantum thermal radiation from the black hole) and its effect on information loss [5-12] have been a serious challenge to modern physics because they require a clear understanding of phenomena ranging from gravity to quantum information theory. Hawking radiation, also called Hawking-Unruh effect, is the origin of the black hole information paradox [5-12]. The problem is that black holes appear to absorb the quantum information as well as the matter, yet the most fundamental laws of physics demand that this information should be preserved as the universe evolves. Key question is whether the pure quantum state decays into mixed states or there is any quantum information left. The essence of Hawking radiation on the scalar particles is that an observer at rest outside a black hole or a uniformly accelerated observer (Rindler spacetime) in the flat spacetime sees a thermal bath of particles [13-15]. On the other hand, Hawking radiation effect on fermions (particles with half-integer spin) and information loss is not so well understood.

In this paper, we report that the Hawking radiation effect on fermions is fundamentally different from the case of scalar particles. Our study shows that the intrinsic properties of fermions (exclusion principle and spin) affect strongly the interaction of fermions with both Hawking radiation and metric of the spacetime. In particular we have found the following: first, while the fermion vacuum state seen by the Rindler observer is an entangled state in which the right and left Rindler wedge states appear in correlated pairs as in the case of the scalar particles, the entanglement disappears in the excited state due to the exclusion principle; second, the spin of the fermion experiences the Winger rotation [16,17] under a uniform acceleration; and third, the quantum information of fermions encoded in the spin degrees of freedom is dissipated not by the Hawking radiation but by the Wigner rotation. As a result, information loss mechanism near the black hole for fermions is quite different from the case of scalar particles. Part of the information survives for scalar particles [18]. For fermions, it is dissipated irreversibly into the curvature of the spacetime.

The basic entity of quantum information is the entanglement [19]. So far, most of the studies on the Hawking radiation and information loss have been done for scalar particles. When entangled scalar particles are under the influence of strong gravitational field of the black hole, its entanglement is affected by the Hawking radiation [16,18,19]. Hawking radiation can be thought as the squeezed state composed of the entangled states inside and outside the black hole [18,20,21]. Information loss near the black hole is related to the mutual information [18]. It was shown that the mutual information converges to unity when one of the entangled pair is absorbed by the black hole [18]. This indicates implicitly that part of the information is hidden in the squeezed vacuum.

*-Hawking-Unruh effect for fermions:* It would be an interesting question to ask whether Hawking radiation also exists for fermions. Recently it was shown that Hawking radiation exists for massive Dirac fermion vacuum state [22-25]. In this paper, we consider entangled fermions in the non-inertial frame and show that the quantum information would be irreversibly lost when it is absorbed by the black hole. The event horizon of the black hole thus acts like a dissipative quantum channel [26,27]. The information is dissipated into the metric of the curved spacetime. Our results are summarized in Table 1.

The essential feature of the Hawking radiation apart from the complications due to the curvature of the spacetime of the black hole is contained in much simpler situation, so called Rindler spacetime [13-15] involving the uniform acceleration of a moving observer Rob and a stationary observer Alice in the flat spacetime. The Schwarzschild spacetime near the event horizon resembles Rindler spacetime in the infinite acceleration limit. Especially, the infinite acceleration limit corresponds to Rob moving on a trajectory arbitrarily close to the Rindler horizon in the context of a black hole and Alice falling into the black hole. Let Alice be an observer at event $P$ with zero velocity in Minkowski spacetime and non-inertial observer Rob be moving with positive uniform acceleration in the x direction with respect to Alice (Fig. 1). Rob is initially moving in the negative direction but comes to stop at event $P$ where Alice is located before moving off in the positive x direction. Rob's trajectory is a hyperbola in the right Rindler wedge labelled region $I$ and bounded by the asymptotes $H_-$ and $H_+$ which represent Rob's past and future horizons. If Rob is under a uniform acceleration, the corresponding vacuum state should be specified in Rindler coordinate in order to describe what Rob observes.

Let's denote the vacuum states for Alice and Rob in Minkowski spacetime as $|O_A\rangle_M$ and $|O_R\rangle_M$, respectively. The Minkowski fermionic vacuum state for Rob can be expressed in terms of Rindler region I and II Fock states as [22-25]

$$|O_R\rangle_M = \frac{1}{(e^{-2\pi\Omega}+1)^{1/2}} \sum_{n=0,1} (-1)^n e^{-n\pi\Omega} |n\rangle_{I\bar{k}\sigma} \otimes |n^c\rangle_{II-\bar{k}\sigma},$$

where $\Omega$ is the energy($\omega / a$), $\bar{k}$ the wave vector, $\sigma$ the spin of the particle. $c$ is the charge conjugation (anti particle), and $|n\rangle_I$ and $|n^c\rangle_{II}$ are the mode decompositions in region $I$ and $II$, respectively. We also define the vacuum states in regions $I$ and $II$ as $|0\rangle_{I\bar{k}\sigma} = |0\rangle_I$ and $|0^c\rangle_{II\bar{k}\sigma} = |0\rangle_{II}$.

The particle creation and annihilation operators for the Rindler space-time are expressed as $b_\eta^\dagger$ and $b_\eta$, respectively. Here, the subscript $\eta = I$ or $II$, takes into account the fact that the space-time has an event horizon, so that it is divided into two causally disconnected Rindler wedges $I$ and $II$ (Fig. 1). The Minkowski operators $a_R^\dagger$ and $a_R$ can be expressed in terms of the Rindler operators $b_\eta^\dagger$ and $b_\eta$ by Bogoliubov transformations [22-25]:

$$a_{R\bar{k}\sigma}^\dagger = \frac{1}{\sqrt{2\cosh\pi\Omega}}\left(e^{\pi\Omega/2}b_{I\bar{k}\sigma}^\dagger - e^{-\pi\Omega/2}b_{II-\bar{k}\sigma}^c\right) \quad a_{R\bar{k}\sigma} = \frac{1}{\sqrt{2\cosh\pi\Omega}}\left(e^{\pi\Omega/2}b_{I\bar{k}\sigma} - e^{-\pi\Omega/2}b_{II-\bar{k}\sigma}^{c\dagger}\right).$$

The state Rob observes is restricted to the right Rindler wedge, i.e., region $I$, in which his motion is confined. Hawking-Unruh effect for the fermionic particles can be expressed as

$$_M\langle O_R | b_{I\bar{k}\sigma}^\dagger b_{I\bar{k}\sigma} | O_R\rangle_M = \frac{1}{1 + e^{2\pi\Omega}},$$

which says a non-inertial observer with uniform acceleration (or an observer near the event horizon of the black hole) would see thermal quantum fields.

Excited state for Rob in Minkowski spacetime is then obtained by applying the Minkowski creation operator $a_R^\dagger$ to the vacuum state:

$$|1_R\rangle_M = a_{R\bar{k}\sigma}^\dagger |O_R\rangle_M$$
$$= \frac{1}{\sqrt{2\cosh\pi\Omega}}\left(e^{\pi\Omega/2}b_{I\bar{k}\sigma}^\dagger - e^{-\pi\Omega/2}b_{II-\bar{k}\sigma}^{c\dagger}\right)\frac{1}{(e^{-2\pi\Omega}+1)^{1/2}}\sum_{n=0,1}(-1)^n e^{-n\pi\Omega}|n\rangle_{I\bar{k}\sigma}\otimes|n^c\rangle_{II-\bar{k}\sigma}.$$
$$= |1\rangle_{I\bar{k}\sigma}\otimes|0\rangle_{II-\bar{k}\sigma} = |1\rangle_{I\bar{k}\sigma}\otimes|0\rangle_{II}$$

Above result shows the fundamental difference of the fermionic vacuum state and the excited state seen by a moving observer Rob. While the ground state is the fermion entangled state in which the right and left Rindler wedge states appear in correlated pairs, the entanglement disappears in the excited state due to the exclusion principle. As a result, Rob in the excited state would not see the quantum bath populated by thermally excited states any more. Very recently, Alsing et al. also reported that the excited state is separable [25]. Their analysis is focused on the entanglement of mode states instead of the spin state entanglement.

-*Wigner rotation in curved space:* So far we have treated the spin of the particle is stationary. In general it is not stationary even if the world line is. The description of spin requires the introduction, at each point, of an independent Lorentz coordinate frame, combined with the demand of invariance under local Lorentz transformations [28]. The relation between the local and the general coordinate system is conveyed by a family of vector fields called tetrads $e_a{}^\mu(x)$, $a = 0, 1, 2, 3$, which respond to general coordinate transformations and local Lorentz transformations as [28] $\bar{e}_a{}^\mu(\bar{x}) = \frac{\partial \bar{x}^\mu}{\partial x^\nu} e_a{}^\nu(x)$ and $\bar{e}_a{}^\mu(x) = \Lambda_a{}^b e_b{}^\mu(x)$, respectively. The metric tensor $g_{\mu\nu}$ is related to the tetrads by $g_{\mu\nu} = e^a{}_\mu e^b{}_\nu \eta_{ab}$, where $\eta_{ab}$ is the metric tensor for Minkowski spacetime. The dual basis $\{\hat{\theta}^a\}$ is defined by $\hat{\theta}^a = e^a{}_\mu dx^\mu$. Then we have $g_{\mu\nu} dx^\mu \otimes dx^\nu = \eta_{ab} \hat{\theta}^a \otimes \hat{\theta}^b$. Throughout the paper, we are using the Latin letters for four local inertial coordinates and the Greek letters for general coordinates. We also note that an infinitesimal displacement in curved spacetime can be viewed as the Lorentz transformation in the local inertial frame [29]. For an infinitesimal coordinate transformation, we get [28]

$$\bar{e}_a{}^\mu(x+\delta x) - \bar{e}_a{}^\mu(x) \rightarrow \delta x^\lambda \frac{\partial}{\partial x^\lambda} e_a{}^\mu(x) + \delta x^\lambda e_a{}^\nu(x) \Gamma^\mu{}_{\nu\lambda}(x) = \delta x^\lambda \nabla_\lambda e_a{}^\mu(x),$$

where $\Gamma^\mu{}_{\nu\lambda}$ is the affine connection and $\nabla_\lambda$ is the covariant derivative.

We also have $\delta x^\lambda \nabla_\lambda e_a{}^\mu = \delta \omega^b{}_a e_b{}^\mu$, with $\delta \omega^b{}_a = \delta x^\nu e^c{}_\nu \Gamma^b{}_{ca}$ and $\Gamma^b{}_{ca} = e^b{}_\kappa e_c{}^\lambda \nabla_\lambda e_a{}^\kappa$. Here, we have used the following orthogonal properties of tetrad $e_a{}^\mu(x)$ and its inverse $e_\mu{}^a(x)$ : $e_\mu{}^a(x) e_\nu{}^a(x) = \delta^\mu_\nu$, $e_a{}^\kappa(x) e_\kappa{}^b(x) = \delta^b_a$. Since any vector V is independent of the basis, it follows that $V^\mu = e_a{}^\mu V^a$ and $V^a = e_\mu{}^a V^\mu$. It is straightforward to show that $\delta \omega_{ba} = -\delta \omega_{ab}$, so the infinitesimal local Lorentz transformation is give by

$$e^a{}_\mu(x) \rightarrow \Lambda^a{}_b(x)\, e^b{}_\mu(x), \quad \Lambda^b{}_a(x) = \delta^b_a - \delta \omega^b{}_a(x),$$

where the connection one form $\delta \omega^a{}_b$ is also obtained from $d\hat{\theta}^a + \delta \omega^a{}_b \wedge \hat{\theta}^b = 0$ [28]. The Hilbert space vector for a spin 1/2 particle is then defined on the local inertial frame spanned by the tetrads. On the tangent plane defined by the basis $\{e_a{}^\mu(x)\}$, an infinitesimal Lorentz transformation $\Lambda$ induced by a coordinate displacement transforms vectors in the Hilbert space by the transformation rule [16,17]

$$|1\rangle_{l\vec{k}\sigma} \rightarrow \sum_{\bar{\sigma}} D^{(j)}_{\bar{\sigma}\sigma}(W(\Lambda, k^a))|1\rangle_{l\Lambda\bar{k}\bar{\sigma}},$$

Here, $W(\Lambda, p)$ is Wigner's little group element, $D^{(j)}(W)$ the representation of $W$ for spin $j$, $k^a = (\vec{k}, k^0)$, $(\Lambda k)^a = (\Lambda \vec{p}, (\Lambda p)^0)$ with $a = 1, 2, 3, 0$, and $L(p)$ is the Lorentz transformation $k^a = L^a{}_b \kappa^b$, and $\kappa^b = (m, 0, 0, 0)$ is the four-momentum taken in the particle's rest frame. The infinitesimal displacement causes the spin of the particle transported to one local inertial frame to another. Throughout the paper we use units $c = G = 1$. Here $c$ is the speed of light. The signature of the metric is defined as $(-, +, +, +)$.

From the metric of the Rindler spacetime [14],

$$ds^2 = -\xi^2 d\eta^2 + d\xi^2 + dy^2 + dz^2 = -\hat{\theta}^0 \wedge \hat{\theta}^0 + \hat{\theta}^1 \wedge \hat{\theta}^1 + \hat{\theta}^2 \wedge \hat{\theta}^2 + \hat{\theta}^3 \wedge \hat{\theta}^3,$$

we obtain the connection one-form $\delta\omega^0{}_1 = -d\eta$, $\eta = a\tau$, $\xi = 1/a$, and the other connection one-forms are zero. Here $\tau$ is the proper time and $d\eta = ad\tau$.

The Wigner matrix is defined by [16,17] $D(W(\Lambda, k)) = D^{-1}(L(\Lambda(k)))D(\Lambda)D(L(k))$. After some mathematical manipulations, we obtain:

$$D(W(\Lambda, k)) = (A\cosh(d\eta/2) + B\sinh(d\eta/2))I - (A\sinh(d\eta/2) + B\cosh(d\eta/2))\sigma_1,$$

where $A = 1 - \dfrac{K^2 d\eta}{(1-K^2)mk^1}$, $B = \dfrac{Kd\eta}{(1-K^2)mk^1}$, $\sigma_1 = \begin{pmatrix} 0 & 1 \\ 1 & 0 \end{pmatrix}$, and $K = \left(\dfrac{k^0 - m}{k^0 + m}\right)^{1/2}$. We assumed that the momentum four vector at the event P is given by $k^0 = m\cosh\delta$, $k^1 = m\sinh\delta$.

Let's assume that at the event $P$, Alice and Rob initially share a maximally entangled state of fermion (spin ½) particles : $\dfrac{1}{\sqrt{2}}\left(|1_A\rangle_{M\vec{k}\sigma_3} \otimes |1\rangle_{l\vec{k}\sigma_3} + |1_A\rangle_{M\vec{k}-\sigma_3} \otimes |1\rangle_{l\vec{k}-\sigma_3}\right)$. Here we consider the entanglement of spin and $\sigma_3$ is the spin in the z direction. The infinitesimal displacement from the event $P$ changes the state into

$$|\Psi_{AR}\rangle = \frac{1}{\sqrt{2}}\left(|1_A\rangle_{M\vec{k}\sigma_3} \otimes D(W(\Lambda, k)|1\rangle_{l\vec{k}\sigma_3} + |1_A\rangle_{M\vec{k}-\sigma_3} \otimes D(W(\Lambda, k)|1\rangle_{l\vec{k}-\sigma_3}\right).$$

**-Fermion entanglement and mutual information in Rindler spacetime:** The density operator seen by Rob in the moving frame is $\rho_{AR} = |\Psi_{AR}\rangle\langle\Psi_{AR}|$. In order to find the entanglement condition, we need to calculate the eigenvalues of the $\rho_{AR}{}^{T_2}$ where $T_2$ is the partial transposition (PT) with respect to Rob. One of the eigenvalues will be negative as long as Alice and Rob is entangled [30]. The measure of the entanglement is an entanglement monotone $E_N$, which is related to the negative eigenvalue $\lambda_-$ by $E_N = 2 \mid \lambda_- \mid \approx \exp\left(-\dfrac{2K^2 d\eta}{(1-K^2)\sinh\delta}\right)$. When both Alice and Rob are stationary (zero acceleration), $E_N$ is equal to one. On the other hand, in the infinite acceleration which corresponds to the case of Alice falling into the black hole, $E_N \rightarrow 0$, asymptotically.

The amount of correlation in the joint state of Alice and Rob is estimated by calculating the mutual information, which measures how much information Alice and Rob have in common [19]. The mutual information is defined by $I(\rho_{AR}) = S(\rho_A) + S(\rho_R) - S(\rho_{AR})$, where $S(\rho) = -tr(\rho \log_2 \rho)$ is the von Neumann entropy. After some mathematical manipulations, we obtain

$$
\begin{aligned}
I(\rho_{AR}) \approx & -\frac{1}{2}\exp\left[-\left(1+\frac{2K}{(1-K)\sinh\delta}\right)d\eta\right]\log_2\left\{\frac{1}{2}\exp\left[-\left(1+\frac{2K}{(1-K)\sinh\delta}\right)d\eta\right]\right\} \\
& -\left\{1-\frac{1}{2}\exp\left[-\left(1+\frac{2K}{(1+K)\sinh\delta}\right)d\eta\right]\right\}\log_2\left\{1-\frac{1}{2}\exp\left[-\left(1+\frac{2K}{(1+K)\sinh\delta}\right)d\eta\right]\right\} \\
& -\frac{2K^2 d\eta}{(1-K^2)\sinh\delta}\exp\left(-\frac{2K^2 d\eta}{(1-K^2)\sinh\delta}\right).
\end{aligned}
$$

When both Alice and Rob are stationary (zero acceleration), $I(\rho_{AR})$ is equal to 2. As the acceleration increases, it decreases and eventually converges to zero in the limit of infinite acceleration. On the other hand, the mutual information of scalar particles is converging to unity [18] when Alice is absorbed by the black hole. Part of Alice-Rob entanglement is shared by the states in region *I* and *II* because Rob's state is composed of squeezed state of region *I* and *II*. However, in the case of fermions, Alice-Rob entanglement is no longer shared with the state in region *II*.

The present results show that the quantum information encoded in the spin degrees of freedom (entangled state is composed of different spin states but with the same mode function) is dissipated not by the Hawking radiation but by the Wigner rotation of spin in curved space. In this work, the effect of final state boundary condition [10-12] of the black hole evolution is not considered and will be left as a future work.

In summary, the effects of Hawking radiation and Wigner rotation on the fermion entangled pair are studied. We found that the Hawking radiation effect on fermions is different from the case of scalar particles. While the fermion vacuum state seen by the Rindler observer is an entangled state in which the right and left Rindler wedge states appear in correlated pairs as in the case of the scalar particles, the entanglement disappears in the excited state due to the exclusion principle. The spin of the fermion experiences the Winger rotation under a uniform acceleration. The quantum information of fermions encoded in spin (entangled state is composed of different spin states but with the same mode function) is dissipated not by the Hawking radiation but by the Wigner rotation as the pair approaches the event horizon..

**Acknowledgements** This work was supported by the Korea Science and Engineering Foundation, the Korean Ministry of Science and Technology through the Creative Research Initiatives Program R-16-1998-009-01001-0(2006). The author thanks M. S. Kim for his help with the figure.

# Table 1 Summary of Hawking radiation effect on scalar particles and fermions.

| | Scalar Particles* | Fermions (this work) |
|---|---|---|
| Vacuum state | $\lvert O_R\rangle_M = \dfrac{1}{\cosh r}\sum\limits_{n=0}^{\infty}\tanh^n r\,\lvert n\rangle_I\otimes\lvert n\rangle_{II}$ <br><br> with $r$ the acceleration parameter defined by $\tanh r = \exp(-2\pi\Omega)$, $\Omega = \lvert k\rvert c/a$, where $k$ is the wave vector, $c$ is the speed of light, and $\lvert n\rangle_I$ and $\lvert n\rangle_{II}$ are the mode decompositions in region $I$ and $II$, respectively. <br><br> Bogoliubov transformation: <br> $a_R^\dagger = b_I^\dagger\cosh r - b_{II}\sinh r$. <br> The particle creation and annihilation operators for the Rindler space-time are expressed as $b_\sigma^\dagger$ and $b_\sigma$, respectively. | $\lvert O_R\rangle_M = \dfrac{1}{(e^{-2\pi\Omega}+1)^{1/2}}\sum\limits_{n=0,1}(-1)^n e^{-n\pi\Omega}\lvert n\rangle_{I\bar k\sigma}\otimes\lvert n^c\rangle_{II-\bar k\sigma}$ <br><br> Here $\Omega$ is the energy, $\bar k$ the wave vector, $\sigma$ the spin of the particle. $c$ is the charge conjugation (anti particle), and $\lvert n\rangle_I$ and $\lvert n^c\rangle_{II}$ are the mode decompositions in region $I$ and $II$, respectively. <br><br> Bogoliubov transformation: <br> $a_{R\bar k\sigma}^\dagger = \dfrac{1}{\sqrt{2\cosh\pi\Omega}}\left(e^{\pi\Omega/2}b_{I\bar k\sigma}^\dagger - e^{-\pi\Omega/2}b_{II-\bar k\sigma}^c\right)$ <br> (ref. 19) |
| Excited state | $\lvert 1_R\rangle_M = a_R^\dagger\lvert O_R\rangle_M$ <br> $=\dfrac{1}{\cosh^2 r}\sum\limits_{n=0}^{\infty}\tanh^n r\sqrt{n+1}\,\lvert n+1\rangle_I\otimes\lvert n\rangle_{II}$ <br><br> $\lvert 2_R\rangle_M = \dfrac{1}{\cosh^3 r}\sum\limits_{n=0}^{\infty}\tanh^n r\sqrt{(n+1)(n+2)}\,\lvert n+2\rangle_I\otimes\lvert n\rangle_{II}$ <br> … <br> Excited states are all entangled. | $\lvert 1_R\rangle_M = a_{R\bar k\sigma}\lvert O_R\rangle_M = \lvert 1\rangle_{I\bar k\sigma}\otimes\lvert 0\rangle_{II}$ <br> Excited state is the product state. |
| Cause of Information loss | Hawking radiation | Wigner rotation of spin |
| Mutual Information at zero acceleration for entangled pair | 2 | 2 |
| Mutual Information at infinite acceleration for entangled pair (Alice is absorbed by the black hole) | 1 | 0 |

* Summarized from refs. 12, 16 and 18.

**Figure legend**

**Figure 1** Rindler spacetime. In region *I* and *II*, time coordinates $\eta =$ constant are straight lines through the origin. Space coordinates $\zeta =$ constant are hyperbolae with null asymptotes $H_+$ and $H_-$, which act as event horizons. The Minkowski coordinates $t, x$ and Rindler coordinates $\eta, \zeta$ are given by $t = a^{-1}\exp(a\zeta)\sinh a\eta$ and $x = a^{-1}\exp(a\zeta)\cosh a\eta$, where $a$ is a uniform acceleration (ref. 11). We consider the case of Alice in stationary and Rob (green hyperbola) under uniform acceleration. Alice and Rob initially share a maximally entangled state of fermion (spin ½) particles at the event *P*.

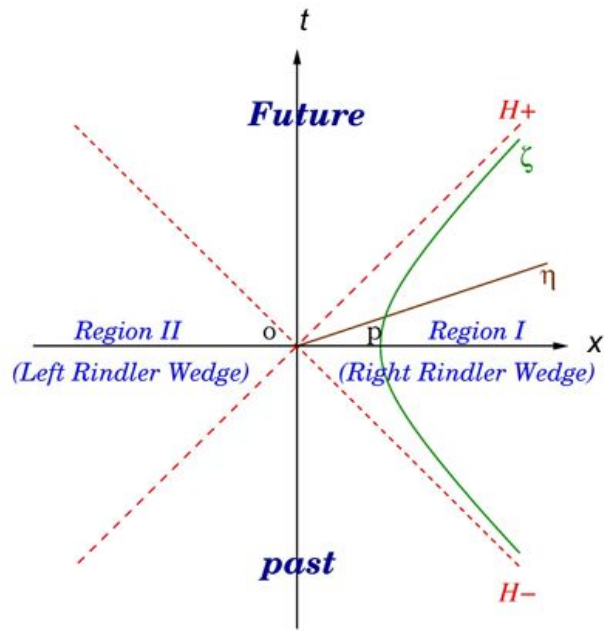

Fig. 1